\documentclass[10pt, conference]{IEEEtran}

\def\BibTeX{{\rm B\kern-.05em{\sc i\kern-.025em b}\kern-.08em
		T\kern-.1667em\lower.7ex\hbox{E}\kern-.125emX}}

\newlength{\bibitemsep}
\newlength{\bibparskip}
\let\oldthebibliography\thebibliography
\renewcommand\thebibliography[1]{%
	\oldthebibliography{#1}%
	\setlength{\parskip}{\bibitemsep}%
	\setlength{\itemsep}{\bibparskip}%
}
\usepackage{amsmath,amssymb,amsfonts}
\usepackage{graphicx}
\usepackage{textcomp}
\usepackage{xcolor}
\usepackage[noadjust]{cite}
\newcommand{\algorithmautorefname}{Algorithm}
\newcommand{\refappendix}[1]{\hyperref[#1]{Appendix~\ref*{#1}}}
\def\sectionautorefname~#1\null{Section #1\null}
\def\equationautorefname~#1\null{Equation (#1)\null}
\def\algorithmautorefname~#1\null{Algorithm (#1)\null}
\def\subsectionautorefname~#1\null{Section #1\null}
\def\subsubsectionautorefname~#1\null{Section #1\null}
\usepackage{tikz}
\usepackage{hyperref}
\definecolor{oceanboatblue}{rgb}{0.0, 0.47, 0.75}
\definecolor{tomato}{rgb}{0.7, 0.0, 0.0}

\hypersetup{
	colorlinks,
	linkcolor={tomato!100!black},
	citecolor={tomato!100!black},
	urlcolor={oceanboatblue!100!black}
}
\usepackage{booktabs}
\usepackage[labelfont=bf, font=footnotesize]{caption}
\usepackage{cleveref}
\usepackage{subcaption}
\usepackage{listings}

\usepackage{enumitem}
\usepackage{ragged2e}
\usepackage{multirow}
\usepackage{arydshln}

\definecolor{lgray}{rgb}{0.98,0.98,0.98}
\definecolor{algray}{rgb}{0.12,0.12,0.12}
\definecolor{algreen}{rgb}{0, 0.5,0.25}
\definecolor{mygray}{rgb}{0.2,0.2,0.2}
\definecolor{mediumgray}{rgb}{0.6, 0.6, 0.6}

\usepackage[ruled, vlined, linesnumbered]{algorithm2e}
\SetKwInOut{Parameter}{parameter}

\SetCommentSty{mycommfont}
\SetAlgorithmName{Algorithm}{Algorithm}
\SetAlCapNameFnt{\footnotesize}
\SetAlCapFnt{\footnotesize}
\usepackage{color,soul}
\usepackage{mdframed}
\usepackage[skins]{tcolorbox}
\tcbuselibrary{breakable}

\definecolor{gostop}{rgb}{0, 0.56, 0.35}
\definecolor{mygreen}{rgb}{0,0.42,0.21}

\lstdefinestyle{str}{
	language=C,
	numbers=none,
	breaklines=true,
	commentstyle=\color{mygray},
	columns=fullflexible,
	basicstyle=\fontsize{8.5}{10}\ttfamily\color{black},
	moredelim=**[is][\bfseries]{@-}{-@},
	moredelim=**[is][\bfseries\color{mygreen}]{@-@}{@-@}, %
	escapeinside={(*}{*)},
	tabsize=4,
	captionpos=t,
	frame=bt,
	framesep=2pt,
	framerule=0.5pt,
	showstringspaces=false
}

\lstdefinestyle{base}{
	language=C,
	numbers=left,
	numberstyle=\scriptsize,
	numbersep=5pt,
	breaklines=true,
	commentstyle=\color{mygray},
	columns=fullflexible,
	basicstyle=\fontsize{7.3}{9}\ttfamily\color{black},
	moredelim=**[is][\bfseries\color{red}]{@-}{-@},
	moredelim=**[is][\bfseries\color{mygreen}]{@-@}{@-@}, %
	escapeinside={(*}{*)},
	tabsize=4,
	xleftmargin=10pt,
	captionpos=t,
	frame=bt,
	framesep=2pt,
	framerule=0.5pt,
	showstringspaces=false
}

\makeatletter
\newcommand{\thickhline}{%
	\noalign {\ifnum 0=`}\fi \hrule height 1.3pt
	\futurelet \reserved@a \@xhline
}
\makeatother

\newcommand{\sys}{\mbox{\textsc{Centris}}\xspace}

\newcommand{\dejavu}{\mbox{D{\'e}j{\`a}Vu}\xspace}

\usepackage{xstring}
\newcommand{\PP}[1]{
	\medskip
	\noindent{\textbf{\IfEndWith{#1}{.}{#1}{#1.}}}}

\newcommand{\PPP}[1]{
	\vspace{1px}
	\emph{\IfEndWith{#1}{.}{#1}{#1.}}
}

\newcommand{\ie}{\textit{i}.\textit{e}.}
\newcommand{\eg}{\textit{e}.\textit{g}.}
\newcommand\ddfrac[2]{\frac{\displaystyle #1}{\displaystyle #2}}

\let\emptyset\O %

\begin{document}

\title{\sys: A Precise and Scalable Approach for Identifying Modified Open-Source Software Reuse}

\author{
	\IEEEauthorblockN{Seunghoon Woo\IEEEauthorrefmark{2}, Sunghan Park\IEEEauthorrefmark{2}, Seulbae Kim\IEEEauthorrefmark{4}, Heejo Lee\IEEEauthorrefmark{2}\IEEEauthorrefmark{1}, Hakjoo Oh\IEEEauthorrefmark{2}}
	\IEEEauthorblockA{\IEEEauthorrefmark{2}\textit{Korea University}, \{seunghoonwoo, sunghan-park, heejo, hakjoo\_oh\}@korea.ac.kr}
	\IEEEauthorblockA{\IEEEauthorrefmark{4}\textit{Georgia Institute of Technology}, seulbae@gatech.edu}
}

\maketitle

\begin{abstract}
	Open-source software (OSS) is widely reused
	as it provides convenience and efficiency
	in software development.
	Despite evident benefits,
	unmanaged OSS components can introduce threats,
	such as vulnerability propagation and license violation.
	Unfortunately, however, 
	identifying reused OSS components 
	is a challenge %
	as the reused OSS is %
	predominantly
	modified and nested. %
	In this paper, we propose \sys,
	a precise and scalable approach for identifying modified OSS reuse.
	By segmenting an OSS code base %
	and detecting %
	the reuse of a unique part of the OSS only, %
	\sys is capable of precisely identifying modified OSS reuse in the presence of nested OSS components.  
	For scalability,
	\sys eliminates redundant code comparisons
	and accelerates the search
	using hash functions.
	When we applied \sys on 10,241 widely-employed GitHub projects,
	comprising 229,326 versions and 80 billion lines of code,
	we observed that modified OSS reuse is a norm in software development, occurring 20 times more frequently than
	exact reuse. %
	Nonetheless,
	\sys identified reused OSS components
	with 91\% precision
	and 94\% recall in 
	less than a minute per application on average,
	whereas a recent clone detection technique, which does not take into account modified and nested OSS reuse, 
	hardly reached 10\% precision
	and 40\% recall. %

\end{abstract}


\section{Introduction}
\smallskip

\newcommand\blfootnote[1]{%
	\begingroup
	\renewcommand\thefootnote{}\footnote{#1}%
	\addtocounter{footnote}{-1}%
	\endgroup
}

Recent years have seen a dramatic surge in the number and use of open-source software (OSS)~\cite{blackduckreport, git100m}.
Not to mention 
the immediate benefit of reusing
the functionalities of existing OSS projects,
using OSS in software development generally leads to improved reliability   
because OSS 
is publicly scrutinized by multiple parties.
At the same time, however,
reusing OSS without
proper management
can  %
impair the maintainability and
security of software
\cite{li2016clorifi, kim2017vuddy, duan2017identifying, kim2018software},
especially when a piece of code is reused over various projects.\blfootnote{*Heejo Lee is the corresponding author.}

One effective solution to prevent this undesirable situation
is to undertake \textit{software composition analysis} (SCA)~\cite{duan2017identifying, sca, gartner}.
The aim of SCA process is to identify the OSS components contained in a target program. 
With an SCA tool,
developers can systematically keep track of what and how OSS components are reused in their software, 
and can therefore mitigate security threats (by patching known vulnerabilities) and avoid potential license violations.

Unfortunately,
precisely identifying OSS components in the target software
is becoming increasingly challenging,  
mainly owing to the following recent trends in software development practice regarding OSS.
\begin{enumerate}
	\item \textbf{Modified OSS reuse:}
	Instead of reusing existing OSS in its entirety,
	developers commonly utilize only a portion of it,
	or modify the source code or structure.
	\item \textbf{Nested OSS components:}
	The reused OSS
	may contain multiple sub-OSS components,
	and even the sub-OSS components
	may include other OSS components.
	
	\item \textbf{Growth of OSS projects and their code size:}
	The number of OSS projects is rapidly increasing~\cite{git100m},
	along with the growing code size~\cite{barb2014statistical}.
\end{enumerate}
These three factors collectively affect the accuracy and scalability of SCA tools.  
To our knowledge, no existing techniques are capable of precise and scalable detection of modified OSS reuse in the presence of nested OSS components.

\PP{Limitations of existing techniques}
Existing SCA techniques %
assume that the reused OSS is essentially unmodified (or modified to a limited fashion), thereby producing false negatives when it comes to identifying modified reuse.  
For example, OSSPolice \cite{duan2017identifying},
a recent SCA technique that aims to identify partially reused components, %
cannot identify OSS components when their directory structures are modified.
On the other hand, 	
existing code clone detection techniques
(\eg, \cite{sajnani2016sourcerercc, lopes2017dejavu, wang2018ccaligner}) %
can, in principle, be used for identifying modified reuse of OSS components, but they easily produce false positives if an OSS project is nested. When only a nested third-party software component of an OSS is used in the target software, clone detection techniques falsely report them as reuse of the original OSS.  
Also, as we demonstrate in this paper, 
existing SCA and clone detection techniques are hardly scalable for large OSS code bases
(details are explained in \autoref{sec:related}).

\PP{Our approach}
In this paper, we present \sys, a new SCA technique that aims to overcome the above limitations. 
\sys~can effectively detect modified OSS reuse in a precise and scalable manner even when OSS components are arbitrarily nested.
For scalability, \sys~uses a technique called {\em redundancy elimination}. 
Instead of generating signatures
from all functions in all versions
of the entire OSS code base,
\sys %
first collects all functions in all versions of an OSS project,
and then removes all redundancies in functions across versions.
This approach is effective in
reducing space complexity;
most of the time,
the delta across versions is significantly smaller
than the size of the unchanged code base.
For precision, we employ a technique called {\em code segmentation}. 
To identify modified components,
we basically use loose matching
that checks whether the code similarity between the target software and the OSS
is greater than a predefined threshold.
However,
simply applying this method %
suffers from false alarms
especially when an OSS is nested.
Therefore,
we segment an OSS into
the \textit{application} code
(\ie, a unique part of the OSS)
and the \textit{borrowed} code
(\eg, a part of the nested third-party software); 
we analyze whether each function in the OSS belongs to 
the application code or the borrowed code.
We then remove the borrowed code of an OSS and only analyze the reuse patterns of 
the application code of the OSS for component identification. 
This code segmentation
enables \sys to drastically
filter out false alarms
while still %
identifying desired OSS components
even when they are heavily modified
or nested.

\PP{Evaluation}
For the experiment,
we collected a large OSS dataset
from
10,241 public C/C++ repositories on GitHub
comprising 229,326 versions and 80 billion lines of code (LoC) in total.
From a cross-comparison experiment,
we discovered that
95\% of the detected components were reused with modifications.
Nevertheless,
\sys successfully identified
the reused components with 91\% precision
and 94\% recall,
whereas a recent clone detection technique,
\dejavu \cite{lopes2017dejavu}, 
yielded less than 10\% precision
and at most 40\% recall
because
\dejavu
neither identifies heavily modified components
nor filters out false alarms caused by nested components
(see \autoref{subsec:acc_others}).
Furthermore,
\sys reduced the matching time
to tens of seconds
when comparing a software with one million LoC
to the dataset, %
while \dejavu %
requires more than three weeks
because they perform matching against 
all lines of code from every OSS in the dataset (see \autoref{subsec:scalability}).

\PP{Contributions}
This paper makes the following contributions:

\smallskip
\begin{itemize}
	\setlength\itemsep{0.5em}
	\item
	We propose \sys, 
	the first approach capable of precisely and scalably identifying
	modified OSS reuse in the presence of nested OSS components.  
	The key enabling technical contributions include redundancy elimination and code segmentation. 

	\item We applied \sys %
	in an industrial setting with a large OSS dataset.
	As a result, we confirmed that 
	most (95\%) of the OSS components are reused with modification.
	\item 
	\sys can identify reused OSS components
	from 10K widely-utilized 
	software projects on GitHub %
	with 91\% precision
	and 94\% recall,
	even though modified OSS reuse is prominent. %
	\sys takes less than a minute on average to identify components
	in a software project.
\end{itemize}
\smallskip

\medskip
\section{Terminology and Motivation}\label{sec:def}
\smallskip

\subsection{Terminology}\label{subsec:termdef}

\PP{Basic terms}
We first define a few terms upfront.
\textit{Target software} denotes the software from which we want to identify reused OSS components.
An \textit{OSS component} refers to an entire OSS package or sometimes the functions 
contained in the OSS; called component for short.
Lastly, \textit{OSS reuse} refers to
utilizing all or some of the OSS functionalities
\cite{krueger1992software, griss1997software}.

\PP{A software project}
We define a software project
as the combined set of
\textit{application} and \textit{borrowed} codes.
The borrowed code denotes the part comprised of reused OSS,
\ie, a set of third-party software,
which we aim to identify within the target software.
The application code refers to the original part of the software project
excluding the code from another OSS.

\PP{OSS reuse patterns}
We classify OSS reuse patterns
into four categories
according to the code and structural changes:

\smallskip
\begin{enumerate}
	\setlength\itemsep{0.2em}
	\item \textbf{Exact reuse (E):}
	The case where
	the entire OSS is reused in the target software
	without any modification.
	\item \textbf{Partial reuse (P):}
	The case where only
	some parts of an OSS
	are reused in the target software.
	\item \textbf{Structure-changed reuse (SC):}
	The case where
	an OSS is reused in the target software
	with structural changes,
	\ie, the name or location of an original file or directory is changed,
	such as code amalgamation.
	\item \textbf{Code-changed reuse (CC):}
	The case where
	an OSS is reused %
	with
	source code changes.
\end{enumerate}
\smallskip
When an OSS is reused with modification
(\ie, partial, structure-changed, and code-changed reuse),
we refer to this as \textit{modified OSS reuse}.
In the modified reuse,
P, SC, and CC can occur simultaneously.

\medskip
\subsection{Motivating example}\label{subsec:problem}

Suppose we want to identify OSS components
reused in ArangoDB v3.1.0 (3.5 million\;LoC),
a native multi-model database software\footnote{\url{https://github.com/arangodb/arangodb}}.
Given a large OSS dataset (80 billion LoC),
\sys took less than a minute
and identified a total of 29 C/C++ OSS components in ArangoDB. %
\autoref{table:arango_res} elaborates on
five of the identified OSS components.

\begin{table}[t]
	\footnotesize
	\renewcommand{\tabcolsep}{0.28mm}
	\begin{center}
		\caption{\label{table:arango_res}Examples of identified components in ArangoDB using \sys.}
		\vspace{-0.5em}
		\begin{tabular}{c|c|c|c|c|c|c}
			\thickhline
			\rule{0in}{2ex}\multirow{2}{*}{\textbf{Name}} 
			& \multirow{2}{*}{\textbf{Version}} 
			& \multicolumn{2}{c|}{\textbf{\#Reused functions}} 
			& \multirow{2}{*}{\begin{tabular}[c]{@{}c@{}}\textbf{\#Unused}\\\textbf{functions}\end{tabular}} 
			& \multirow{2}{*}{\begin{tabular}[c]{@{}c@{}}\textbf{Structure}\\\textbf{change}\end{tabular}} 
			& \multirow{2}{*}{\begin{tabular}[c]{@{}c@{}}\textbf{Reuse}\\\textbf{patterns$^\dagger$}\end{tabular}} 
			\\\cline{3-4}
			&& \rule{0in}{2ex}{Identical} & {Modified} &&&\\\hline
			\rule{0in}{2ex}Curl&v7.50.3&2,211&\textcolor{tomato}{26}&1&\textcolor{black}{\text{\sffamily X}}&P \& CC\\
			\rule{0in}{1.5ex}GoogleTest&v1.7.0&1,197&\textcolor{tomato}{11}&33&\textcolor{tomato}{\text{\sffamily O}}&P \& SC \& CC\\
			\rule{0in}{1.5ex}Asio&v1.61.0&941&\textcolor{tomato}{0}&0&\textcolor{black}{\text{\sffamily X}}&E\\
			\rule{0in}{1.5ex}Velocypack&OLD$^\ddagger$&134&\textcolor{tomato}{0}&3,765&\textcolor{black}{\text{\sffamily X}}&P\\
			\rule{0in}{1.5ex}TZ&v2014b&89&\textcolor{tomato}{0}&26&\textcolor{black}{\text{\sffamily X}}&P\\
			\thickhline
		\end{tabular}

		\scriptsize{\vspace{0.3em}$^\dagger$E: Exact reuse, P: Partial reuse, SC: Structure-changed reuse, CC: Code-changed reuse\\$^\ddagger$OLD version: Velocypack code committed in 2016.}
		
		\vspace{-2.2em}
	\end{center}
\end{table}

The modified reuse pattern is very prominent in ArangoDB.
Among the 29 identified OSS components,
22 were modified, wherein
the reused functions were located in directories different from those in the
original OSS, \eg, GoogleTest,
or the code base was partially updated, \eg, Curl.
Also in most cases, 
ArangoDB reused a fraction of the OSS code base,
\eg, 3.6\% of Velocypack,
with unnecessary features such as testing infrastructure removed.
Moreover, 21 components were reused in the form of nested components;
for example, TZ was reused by the V8 engine, and V8 was in turn reused in ArangoDB.
Existing SCA techniques %
are not designed for handling such code bases with %
modified components.
For example,
six components were reused in ArangoDB with structural changes,
as OSSPolice \cite{duan2017identifying}
relies on the original OSS structure for component detection,
it fails to identify such structure-changed components.
In contrast,
code clone detection techniques
report numerous false alarms
in identifying modified and nested components.
For instance,
\dejavu \cite{lopes2017dejavu}
reported that 422 OSS were reused in ArangoDB,
among which 411 were confirmed as false alarms as we investigated
(see \autoref{subsec:acc_others}).
This is because
\dejavu reports any OSS as a reused component
if the OSS contains
the same third-party software reused in ArangoDB.
One example %
is Ripple, a cryptocurrency-related
OSS that contains RocksDB as a sub-component.
ArangoDB also reuses multiple functions from RocksDB,
thereby having shared functions with Ripple,
and \dejavu misinterprets this relation
as ArangoDB reusing Ripple.

\begin{figure}[t]
	\begin{center}
		\includegraphics[width=1\linewidth]{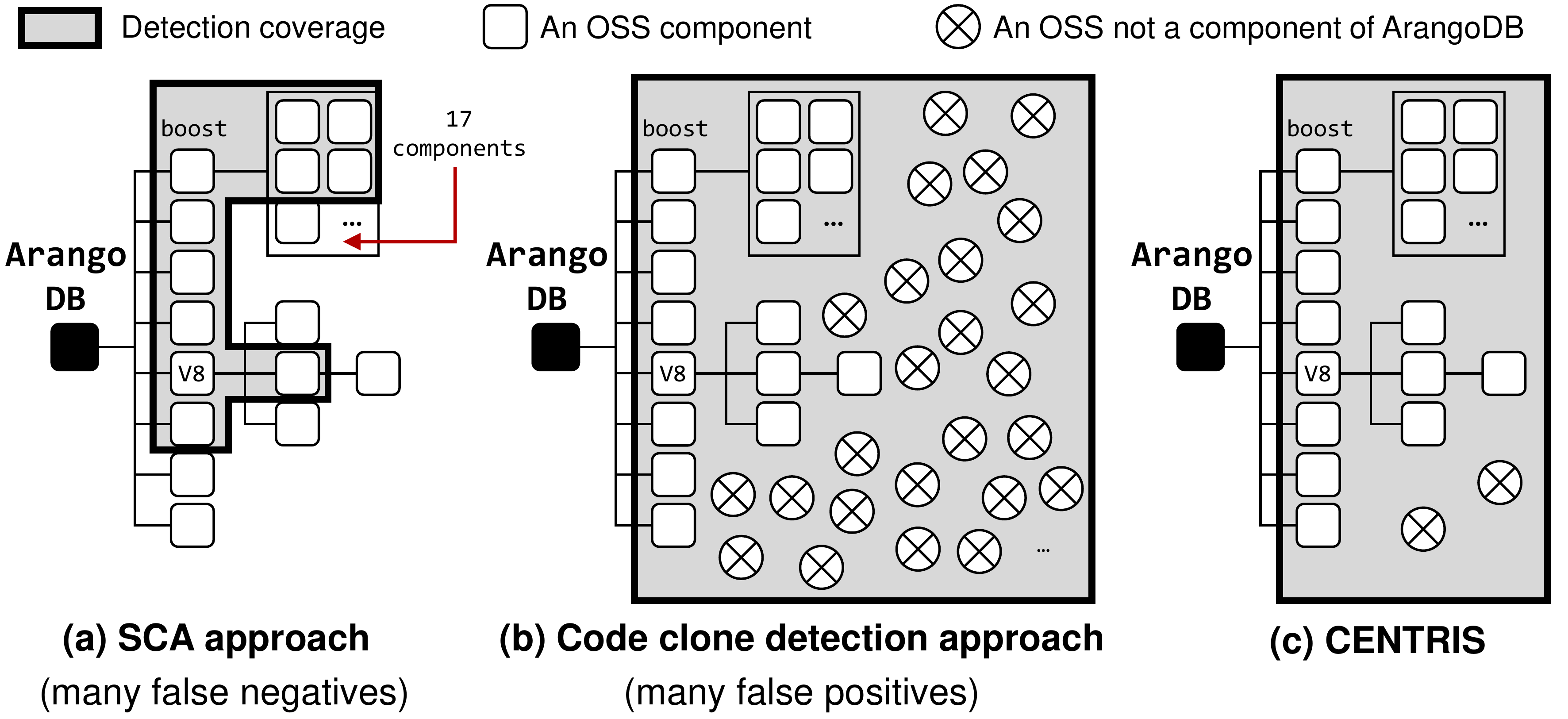}
		\vspace{-0.6em}
		\caption{\label{fig:arango_coverage} Illustration of the component detection coverage of the
			SCA approach, code clone detection approach, and \sys.
			Compared to \sys, which identified 29 components,
			existing SCA approaches could not detect components where structural modification occurs (\eg, OSSPolice~\cite{duan2017identifying})
			or when only a small portion of an OSS code base is reused,
			whereas code clone detection approaches (\eg, \dejavu~\cite{lopes2017dejavu}) reported numerous false positives.
		}
	\end{center}

	\vspace{-0.3em}
\end{figure}

\medskip
\section{Design of \sys}\label{sec:design}

In this section,
we describe the %
design of \sys. %

\subsection{Overview}
\smallskip

\begin{figure}[t]
	\begin{center}
		\includegraphics[width=1\linewidth]{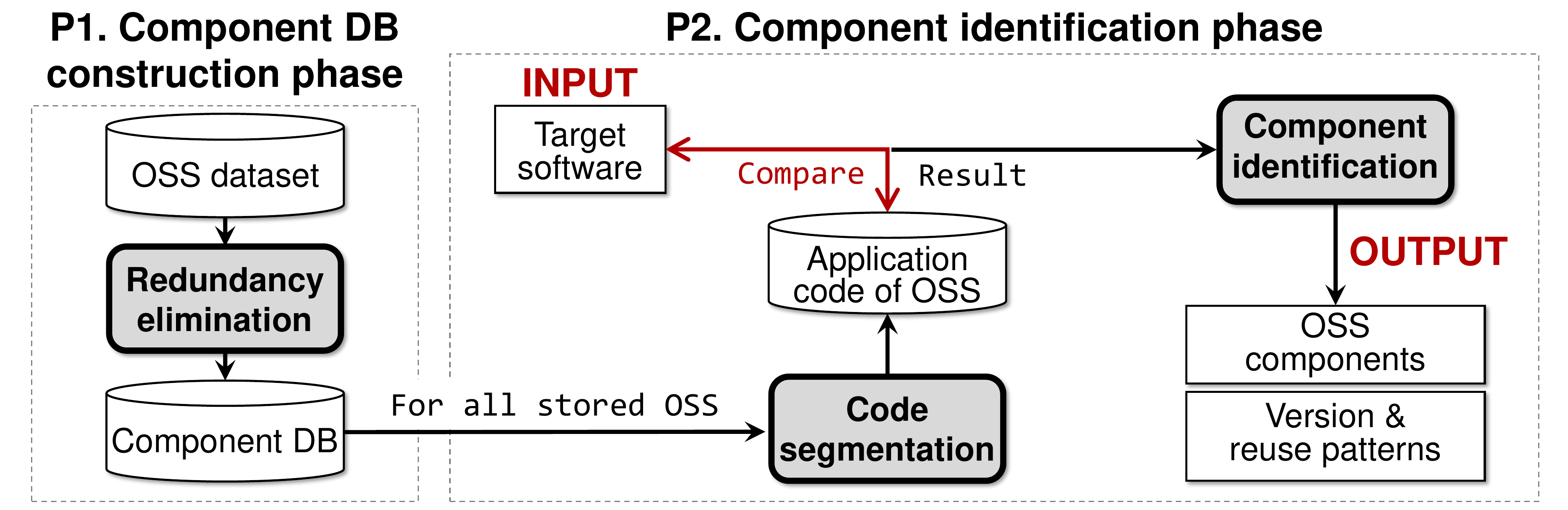}
		\caption{\label{fig:over}
			High-level overview of the workflow of \sys.
		}
	\end{center}
	\vspace{-0.4em}
\end{figure}

\autoref{fig:over} depicts the workflow of \sys.
\sys comprises two phases:
(1) \textbf{P1} for constructing the OSS component database (DB), 
and (2) \textbf{P2} for identifying OSS components reused in the target software. 
In \textbf{P1}, %
we use a technique, called redundancy elimination,
which enables scalable component identification; 
\sys reduces
the space complexity in component identification
by eliminating redundancies across the versions of each OSS project.
All functions of an OSS project are converted into the \textit{OSS signature},
which is a set of functions without redundancies, and subsequently stored in the component DB.
In \textbf{P2}, %
we use a technique, code segmentation, for precise component detection.
Specifically,
\sys minimizes false alarms in component detection 
by only analyzing the patterns wherein the application code
of an OSS is reused in the target software.

\PP{Design assumptions}
\sys is designed to identify OSS components at the source code level;
that is, 
our goal is to 
identify components regardless of whether all or only parts of the OSS code base
are reused in the target software.
In addition,
although the concept of \sys is applicable to any granularity of component units, 
we focus on the \textit{function} units
for the approach design and evaluation.
As the term ``OSS reuse'' refers to
utilizing all or some OSS functionalities
\cite{krueger1992software, griss1997software, duan2017identifying},
we determined that function units %
are more appropriate for detecting various OSS reuse patterns compared to other units.
With less granularity (\eg, a file),
\sys can identify components faster than when using function units, however, \sys may miss partial reuses especially when only some functions in a file were reused in the target software
(the benefits of function units %
have been discussed in previous studies~\cite{sajnani2016sourcerercc, kim2017vuddy, duan2017identifying, duan2019automating}).
In light of this,
\sys extracts functions
from all versions of the OSS in our dataset
using a function parser (see \autoref{sec:imple}),
and performs lightweight text preprocessing
to normalize the function
by removing 
comments, tabs, linefeed, and whitespaces, which 
are easy to change but do not affect program semantics. %

\smallskip
\subsection{Component DB construction phase (P1)}\label{subsec:p1}
\smallskip
In this phase,
we process the OSS projects to generate the component DB.
However,
we observed that simply storing all functions from
all versions of every OSS makes the component identification phase %
extremely inefficient. 

\PP{Redundancy elimination}
We thus focus on the characteristics of OSS:
the entire source code of an OSS is not newly developed each time the OSS is updated,
and thus some parts common to different versions are
redundantly compared with the target software when identifying OSS components.
This characteristic gives the following intuition:
if the functions common to multiple versions
are only once compared with the target software,
space and time complexity can be reduced.

Let us define an OSS signature as
a set of functions of the OSS,	
which will be stored in the component DB.
The process for generating an OSS signature is as follows:

\smallskip
\begin{enumerate}
	\setlength\itemsep{0.42em}
	\item First, we extract all functions in all versions of an OSS.
	\item Next, we create as many bins as the total number of versions in the OSS (denoted as $n$).
	\item When a particular function appears in $i$ different versions of the OSS,
	the function is stored in the $i$-th bin,
	along with the version information to which this function belongs,
	and the path information within each version.
\end{enumerate}
\smallskip

\begin{figure}[t]
	\begin{subfigure}[b]{\linewidth}
		\includegraphics[width=0.85\linewidth]{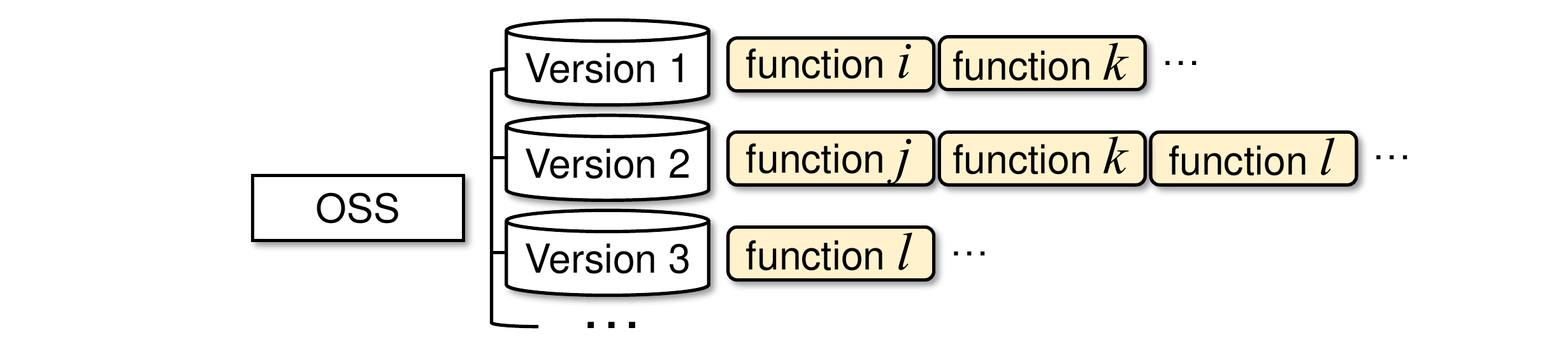}
		\caption{\label{subfig:duple}A naively generated OSS signature.}
	\end{subfigure}
	\begin{subfigure}[b]{\linewidth}
		\includegraphics[width=0.9\linewidth]{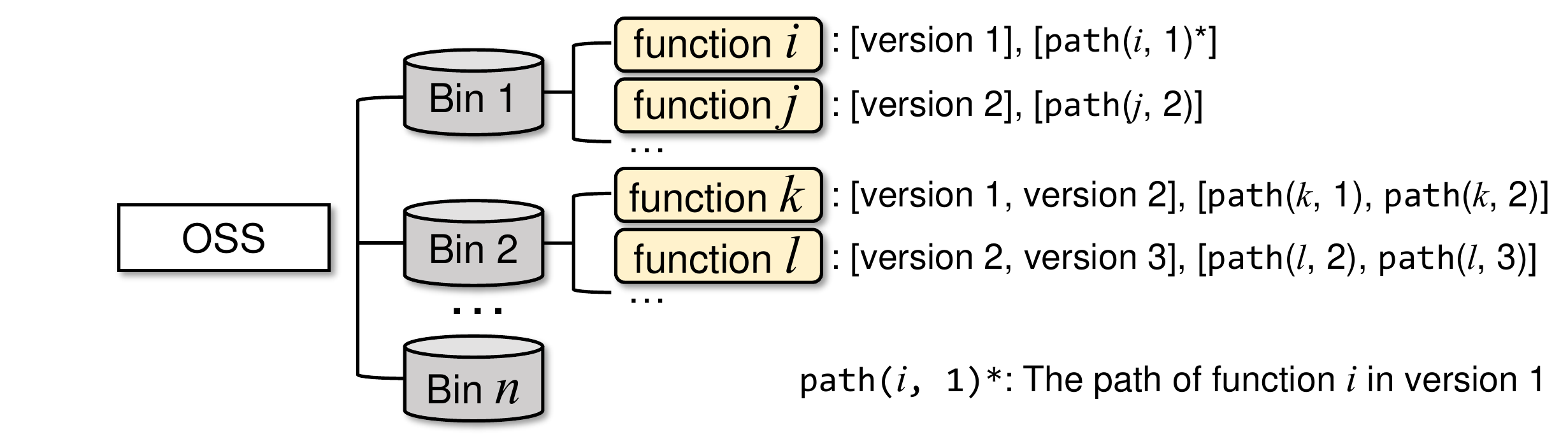}
		\vspace{0.05em}
		\caption{\label{subfig:deduple}A redundancy eliminated signature for an OSS.}	
	\end{subfigure}
	\caption{\label{fig:comp_indexing} Illustration of OSS signatures. We generate signatures for each OSS in the manner shown in (b), thereby reducing space complexity.}
\end{figure}

Note that
all the functions have undergone text preprocessing
in accordance with our design assumptions.
In addition,
we %
apply
a Locality Sensitive Hash (LSH) to the functions
when storing them,
which has native support for measuring the similarity between
two hashes.
The generated $n$ bins of an OSS
become the signature of the OSS
(see \autoref{subfig:deduple}).

If we naively generate a signature by mapping the function to the version it belongs to
(see \autoref{subfig:duple}),
a function that exists in $i$ different versions
would be compared $i$ times with the target software.
However, our method of storing redundant functions only once in the corresponding bin
reduces such unnecessary comparisons;
the quantitative efficiency of
redundancy elimination
is described in \autoref{subsec:scalability}.
Another advantage is that even if an OSS is constantly updated,
the number of functions newly added to the component DB is not large enough to impair scalability.
Lastly,
because there are no
functions excluded from indexing,
if we designed
an appropriate identification algorithm,
the accuracy and, 
specifically,
recall would not be impaired.
By generating signatures for all OSS and storing them, 
the component DB is constructed.

\medskip
\subsection{Component identification phase (P2)}\label{subsec:p2}
\smallskip

In this phase,
\sys identifies the reused OSS components
in the target software.

\PP{Common functions}
We first define the notion of common functions between two software projects.
Each LSH algorithm provides its own \textit{comparison method} and \textit{cutoff} value \cite{lee2017comparison}. 
Using the comparison method,
we can measure the \textit{distance} for each function pair
between the two software projects,
which indicates the syntactic difference
between the two input functions.
Hence,
we define the relation between two functions ($f_1, f_2$)
based on the \textit{distance} and \textit{cutoff} as follows:

\smallskip
\begin{center}
	\noindent\fbox{%
		\parbox{0.97\linewidth}{%
			\begin{itemize}[leftmargin=1.3em]
				\setlength\itemsep{0.2em}
				\item[] \textbf{\textit{LSH-based function relation decision:}}
				\item If $\big($\textit{distance}($f_1, f_2$)$\;=0\big)$: $f_1$ and $f_2$ are \textbf{identical};
				\item {If} $\big(0 <$\textit{distance}($f_1, f_2$)$\;\leq\;$\textit{cutoff}$\big)$: $f_1$ and $f_2$ are {\textbf{similar}};
				\item {If} $\big($\textit{distance}($f_1, f_2$)$\;>\;$\textit{cutoff}$\big)$: $f_1$ and $f_2$ are {\textbf{different}}.
			\end{itemize}
		}%
	}
\end{center}
\smallskip
The similar and identical function pairs
between the two software projects
are determined as the common functions
(the LSH algorithm is specified in \autoref{sec:imple}).

\PP{Key concepts for precise identification}
To identify modified components,
we employ similarity threshold-based loose matching,
\ie, %
to check whether the code similarity between the target software and the OSS is greater than the predefined threshold.
However,
as previously mentioned,
a simple threshold-based identification method %
suffers 
from a large number of false alarms.
False alarms may occur when (i) 
an OSS is nested %
or (ii) 
only the \textit{borrowed code} of the OSS
is included in the target software.
Consequently,
we present
two concepts to reduce false alarms and precisely identify OSS components.

\smallskip
\begin{itemize}
	\setlength\itemsep{0.36em}
	\item \textbf{Prime OSS.}
	This refers to an OSS not containing any third-party software.
	If there is a number of common functions between a prime OSS and the target software,
	the prime OSS can be considered the correct component, 
	because it violates condition (i) for false alarms.
	\item \textbf{Code segmentation.}
	If we only consider the \textit{application code}
	of an OSS in component identification, 
	no false alarms occur owing to the third-party software
	because this does not satisfy the false alarm condition (ii).
\end{itemize}
\smallskip

Accordingly,
our component identification process comprises the following three steps (S1 to S3):

\medskip
\begin{itemize}
	\setlength\itemsep{0.3em}
	\item[\textbf{S1)}] Detecting the prime OSS in the component DB;
	\item[\textbf{S2)}] Extracting application code from all OSS projects;
	\item[\textbf{S3)}] Identifying components within the target software.
\end{itemize}
\medskip

The above steps are conducted
after extracting all functions of the target software
and then applying the text preprocessing and LSH algorithm to the extracted functions.

\medskip
\subsubsection {\textbf{Detecting the prime OSS in the component DB}}
Let $S$ be the OSS to be checked as to whether it contains any third-party software.
To determine whether $S$ is the prime OSS,
we first detect common functions between
$S$ and
each OSS (denoted as $X$) 
in the component DB.
If there is an OSS project having one or more common functions with $S$,
the relation between $S$ and $X$ can be determined as belonging to one of the following four categories
({R$\boldsymbol{_1}$} to {R$\boldsymbol{_4}$}, see \autoref{table:relationships}).

\begin{table}[h]
	\footnotesize
	\renewcommand{\tabcolsep}{3mm}
	\begin{center}
		\vspace{1.5em}
		\caption{\label{table:relationships}
			Possible relations between $\boldsymbol{S}$ and $\boldsymbol{X}$
		}
		\vspace{-0.5em}
		\begin{tabular}{cl}
			\thickhline
			\rule{0in}{2.2ex}\textbf{Type} & \textbf{Description}\\\hline
			\rule{0in}{2.5ex}\textbf{R$\boldsymbol{_1}$.}&$S$ and $X$ share widely utilized codes (\eg, hash function).\\ \hdashline[2pt/2pt]
			\rule{0in}{2.5ex}\textbf{R$\boldsymbol{_2}$.}&$S$ and $X$ simultaneously reuse some other OSS projects.\\
			\hdashline[2pt/2pt]
			\rule{0in}{2.5ex}\textbf{R$\boldsymbol{_3}$.}&$S$ reuses $X$.\\
			\hdashline[2pt/2pt]
			\rule{0in}{2.5ex}\textbf{R$\boldsymbol{_4}$.}&$X$ reuses $S$\\
			\thickhline
		\end{tabular}
	\end{center}	
\end{table}
\noindent Among these relations,
we are interested in {R$\boldsymbol{_2}$} and {R$\boldsymbol{_3}$}, which imply that 
$S$ contains at least one third-party software;
conversely, when $S$ and every $X$ are related by 
{R$\boldsymbol{_1}$} or {R$\boldsymbol{_4}$}, 
we can determine that $S$ is the prime OSS.

In fact,
{R$\boldsymbol{_1}$} contrasts with the other three relations
because there are few common functions between $S$ and $X$.
Therefore,
the main challenge
in determining whether $S$ is the prime OSS
is to differentiate {R$\boldsymbol{_4}$} from {R$\boldsymbol{_2}$} and {R$\boldsymbol{_3}$}.

Subsequently,
we focus on when a common function between $S$ and $X$ first appeared in each OSS;
we refer to this as the \textit{birth time} of the function.
Suppose that $X$ reuses $S$ (\ie, {R$\boldsymbol{_4}$});
then, the birth time
of a particular reused function $f$ in $S$
would be earlier than that in $X$.

Based on the above idea,
we calculate the similarity score ($\phi$)
between $S$ and $X$ as follows
(let $birth (f, S)$ be
the birth time of $f$ in $S$):
\begin{center}
	\vspace{-0.5em}
	$\phi(S, X) = \ddfrac{|G|}{|X|},$\\
	\vspace{0.8em}
	\small{where $G = \{ f\;|\; \big(f \in (S\cap X)\big) \wedge	\big(birth(f, X) \leq birth(f, S)\big) \}$}		
\end{center}

As shown in the above equation,
we measure the similarity score
by considering only the common functions that appeared earlier in $X$ than $S$
for identifying that $X$ exhibits the {R$\boldsymbol{_2}$} and {R$\boldsymbol{_3}$}
relations with $S$.
As there are several ways to obtain
the birth time of a function in an OSS,
\eg, code generation time,
we utilize the information we already have.
Within a bin of an OSS signature,
the function hash values and version information to which the functions belong to are recorded.
Therefore,
we assign
the release date of the earliest version among all recorded versions of a function
as the birth time of the function in the OSS.
In addition,
widely used generic code,
\eg, hash functions or error-handling routines,
can exist in both $S$ and $X$ ({{R$\boldsymbol{_1}$}}),
and thus,
we use $\theta$ as a threshold.

Finally,
we determine that $X$ belongs to the {R$\boldsymbol{_2}$} or {R$\boldsymbol{_3}$} relation if $X$ satisfies the following condition: %
\begin{equation}\label{eq:reuse_condition}
\phi(S,X) \geq \theta
\end{equation}

One might consider that
$X$ could reuse a third-party software (denoted as $R$) at a time later than $S$.
In this case,
because the functions in $R$ have earlier birth times in $S$
than those in $X$,
the functions would not affect the measurement of $\phi(S, X)$.
Therefore, even though $S$ and $X$ contain common third-party software,
\autoref{eq:reuse_condition} may not be satisfied.
However,
this case has no effect on determining whether $S$ is the prime OSS.
Obviously, $S$ contains the $R$ code base,
and even if $\phi(S,X)$ does not satisfy \autoref{eq:reuse_condition},
$\phi(S,R)$ will be greater than $\theta$;
and thus, $S$ is not the prime OSS,
which is the correct answer.

Consequently,
if there is no $X$
that satisfies \autoref{eq:reuse_condition},
we determine that $S$ is the prime OSS.
\begin{equation*}
S =
\begin{cases}
\text{Prime OSS} &  \text{if $\forall X \boldsymbol{.} \big(\phi(S,X) < \theta\big)$}\\
\text{Non-prime OSS} & \text{if $\exists X \boldsymbol{.} \big(\phi(S,X) \geq \theta\big)$}\\
\end{cases}       
\end{equation*}
Otherwise,
we consider every $X$ that satisfies \autoref{eq:reuse_condition}
as possible members of $S$, and store them;
this information will be utilized for the code segmentation.

\vspace{0.73em}
\subsubsection {\textbf{Extracting application code}}

In this step,
we extract the application code through code segmentation for every OSS
in the component DB.
As a prime OSS does not have any borrowed code,
we only focus on non-prime OSS projects.
Let $S$ be the OSS of interest
(\ie, the signature).
One way to locate the application code of $S$ ($S_A$)
is to remove the borrowed code ($S_B$) from $S$
(\ie, $S_A = S \setminus S_B$).
However, 
detecting the OSS that belongs to $S_B$ 
leads to a paradox:
the \sys methodology for identifying
OSS components from the target software
requires the same methodology
for identifying the components of an OSS. %

Fortunately,
we do not need to exactly identify
the sub-components of $S$.
Instead,
we use the possible members of $S$ (denoted as $P$),
which were obtained from the previous step.
As $P$ is a possible member of $S$,
it would be reused in $S$ (\ie, {R$\boldsymbol{_2}$}) or
it reuses a common third-party software with $S$ (\ie, {R$\boldsymbol{_3}$}).
$P$ has no code that might belong to the application code of $S$;
this is because only the code of an OSS that exhibits the {R$\boldsymbol{_4}$} relation with $S$
can belong to the application code of $S$. 
In other words,
the common functions between $S$ and $P$ are
exactly included in the borrowed code of $S$,
as mentioned in our definition (see \autoref{subsec:termdef}).

Therefore,
we can obtain the application code of $S$
by removing all functions of the possible members of $S$ from the function set of $S$.
The high-level algorithm for code segmentation
is shown in \autoref{alg:segmentation}.
Consequently, every OSS project in the component DB remains in a state wherein it is
(i) detected as the prime or
(ii) the application code is extracted (only for the non-prime OSS projects).

\newlength{\commentWidth}
\setlength{\commentWidth}{7cm}
\newcommand{\atcp}[1]{\tcp*[r]{\makebox[\commentWidth]{#1\hfill}}}

\begin{algorithm}[t]
	\caption{\footnotesize \label{alg:segmentation}The high-level algorithm for the code segmentation}\small
	\DontPrintSemicolon
	\KwIn{$S\quad\;$ \tcp*[h]{The OSS that will be segmented}}
	\KwIn{$DB\;$ \tcp*[h]{The component DB}}
	\KwOut{$S_A$ \tcp*[h]{The application code of the $S$}}
	\SetInd{0.8em}{0.8em}
	\SetKwBlock{Begine}{procedure}{end procedure}
	\BlankLine
	\Begine($\textcolor{tomato}{\text{\textsc{\textbf{codeSegmentation}}}} {\big(}S, DB{\big)}$){
		{$S_A \gets \emptyset$\\}
		{$isPrime, members \gets$ \textcolor{black}{\textsc{\textbf{checkPrime}}}$(S, DB)$\\}
		\If{$\neg$($isPrime$) \tcp*[r]{$S$ has borrowed code parts}}
		{
			\For{$P \in members$}
			{
				$S = (S \setminus P)$ \tcp*[r]{Set minus operation}
			}
		}
		{$S_A = S$\\}
		\Return{$S_A$}
	}
	\Begine($\textcolor{black}{\text{\textsc{\textbf{checkPrime}}}} {\big(}S, DB{\big)}$){
		{$isPrime\gets $\texttt{True}\\}
		{$members\gets \emptyset$\\}
		\For{$X \in DB$}
		{
			{$G \gets \emptyset$\\}	
			\For{$f \in (S\cap X)$}
			{
				\If{$birth(f, X) \leq birth(f,S)$}
				{
					{$G.add(f)$\\}
				}					
			}				
			{$\phi (S, X) = (|G| / |X|)$\\}
			\tcp*[r]{Similarity measurement}
			
			\If{$\phi (S, X) \geq \theta$}
			{
				{$isPrime \gets$\texttt{False}\\}
				{$members.add(X)$}
			}
		}
		\Return{$isPrime, members$}
	}
\end{algorithm}
\normalsize

\medskip
\subsubsection {\textbf{Identifying components}}
The next step is identifying the OSS components of the target software. %
Let $T$ be the target software
and $S$ be the OSS in the component DB. %
To identify
whether $S$ is the correct component of $T$,
we measure the code similarity score
between $T$ and the application code of $S$.
If $S$ is the prime OSS,
the application code ($S_A$)
is the same as the entire $S$.
The similarity score ($\Phi$) is calculated
as follows:
\vspace{-0.5em}
\begin{center}
	$\Phi(T, S) = \displaystyle\frac{|T \cap S_A|}{|S_A|}$
\end{center}   
\vspace{0.2em}

There may be a possibility that
widely used or generic code exists in both $T$ and $S$,
as in the case of {R$\boldsymbol{_1}$};
thus, we again employ the threshold $\theta$ as a filter.
Finally,
we determine that $S$ is the correct component
when $\boldsymbol{\Phi(T, S) \geq \theta}$.
Once this process has been applied to all OSS in the component DB,
we can get a set of OSS components of the target software.

\vspace{0.9em}
\noindent\textbf{Why \sys is accurate.}
First,
as \sys does not rely on
structural information in the identification phase,
we can identify components
regardless of
structural change.
Next,
irrespective of
whether OSS is nested,
if the ratio of application code of each OSS
is reused greater than $\theta$,
it can be identified as a correct component.
Lastly,
the code segmentation of \sys
not only reduces false positives,
but also helps to identify heavily modified components.
Let consider the Velocypack component of ArangoDB, introduced in \autoref{subsec:problem};
only 3.6\% of Velocypack code base were reused in ArangoDB.
In fact, Velocypack included another OSS (GoogleTest),
and the ratio of the reused application code of Velocypack was measured as 12\%.
Highlighting the reuse patterns of only the application code of an OSS
makes the similarity score between the target software and the OSS
higher if the OSS is the correct component,
and 
lower when the OSS is a false positive (\ie, close to 0\%).
Using this distinct similarity score difference,
\sys can precisely identify modified components
with low false positives.

\vspace{0.2em}
\PP{Version identification}
To identify the reused version of each component,
we focus on the reused functions of the OSS component. %
In the modified reuse,
the functions of multiple versions
could be simultaneously reused in the target software.	
Therefore, 
we assign a \textit{weight} to each reused function.
Specifically,
we utilize a weighting algorithm
that satisfies the condition
that a larger weight is assigned to functions
belonging to fewer versions.
TF-IDF
\cite{salton1986introduction}
suffices,
where Term Frequency (TF)
refers to the frequency of a function appearing in a particular version
and Inverse Document Frequency (IDF)
refers to the inverse of the number of versions
containing this function.
The IDF that satisfies the condition we set
is utilized as the main weight function,
and %
we use the ``Boolean Frequency'' as the TF;
\ie, we assign 1 to the TF of
all functions.

Let $n$ be the total number of versions of an OSS,
and $V(f)$ be the versions to which a particular function $f$ belongs.
The weight function $W$
is defined as $W(f) = \log{\big(n / |V(f)|\big)}$.
Note that the $|V(f)|$ value of $f$ that belongs to the $i$-th bin is $i$,
by the definition of our signature generation.
Accordingly,
we %
loop through all the reused functions
of the OSS component
and add the weight of each function
to the score of the versions
to which it belongs.
After scoring all functions,
we identify the utilized version
with the highest score.

\vspace{0.1em}
\PP{Reuse pattern analysis}
We then analyze the reuse pattern of the detected components.
First,
to identify code changes occurring during OSS reuse,
we utilize the \textit{distance}
measured using the comparison method of the LSH algorithm
(as explained at the beginning of P2)
for each function pair between the OSS component and the target software.
We determine whether the function is
reused (\textit{distance} = 0),
not reused (\textit{distance} $>$ \textit{cutoff}),
or reused with code changes ($0 < $ \textit{distance} $\leq$ \textit{cutoff}).
Next,
to measure the structural changes,
we analyze the path differences between the reused functions and original functions.
We split each
function path using ``$\,$/$\,$'' (slash),
traverse each path backward
starting from the filename, and compare each path level.
The criterion for comparison is the path of the original function.
If any directory or file name is different,
we determine that the structure has been modified.

Finally,
according to the definition in \autoref{subsec:termdef},
if all functions of the OSS are reused without any modification,
we refer to it as \textit{exact reuse}.
If there are unused functions,
we refer to it as \textit{partial reuse}.
If %
the structure is changed while reusing,
\textit{structure-changed reuse} occurs.
If any code is modified while reusing the OSS,
we refer to it as \textit{code-changed reuse}.

\medskip
\section{Implementation of \sys}\label{sec:imple}
\smallskip

\sys comprises three modules:
an \textit{OSS collector},
a \textit{preprocessor}, and
a \textit{component detector}. %
The OSS collector gathers the source code of popular OSS projects.
The preprocessor stores the OSS signatures generated through redundancy elimination,	
and then extracts the application code of the OSS through code segmentation.
The OSS collector and	
preprocessor
need to be executed only once.
Thereafter,
the component detector %
performs the actual component identification
on the target software.
\sys is implemented in approximately 1,000 lines of Python code, excluding external libraries.

\PP{Initial dataset}
Many programming languages provide
dependency information (\eg, Gemfile in Ruby).
However, C and C++,
which are two of the most popular languages
(combined rank 3 in GitHub), %
do not provide dependency information
despite the need.
Although \sys is not restricted to a particular language,
we demonstrate %
\sys targeting C/C++ software
to prove its efficiency
without any prior knowledge of dependency.
We targeted GitHub,
which has the largest number of repositories
among version control systems
\cite{whygithub}.
Finally,
we %
collected all repositories
having more than 100 stars. %
The OSS collector of \sys
collected 10,241 repositories including
Linux kernel, OpenSSL, Tensorflow, among others (as of April. 2020).
When we extracted all versions (\eg, tags) %
from the repositories,
we obtained 229,326 versions;
the total lines were 80,388,355,577.
This dataset is significantly larger
than those used in previous approaches
(\eg, 2.5 billion C/C++ LoC database~\cite{duan2017identifying}).

\PP{Parser and LSH algorithm}
To extract functions from software,
we employed universal Ctags \cite{ctags},
an accurate and fast open-source regular expression-based parser.
Next,
among the various LSH algorithms
\cite{kornblum2006identifying, roussev2009hashing, oliver2013tlsh},
we selected the TLSH, %
as it is known to incur fewer false positives,
and have a reasonable hashing and comparison speed
as well as low influence of the input size~\cite{lee2017comparison, oliver2013tlsh}.
Its comparison algorithm,
\texttt{diffxlen},
returns the quantified \textit{distance}
between the two TLSH hashes
provided as inputs.
In the context of \sys,
functions that undergo modification after reuse
fall into this category.
We set the cutoff value to 30, referring to \cite{oliver2013tlsh}.

\medskip
\section{Evaluation and Findings}\label{sec:eval}
\smallskip

In this section,
we evaluate \sys. 
\autoref{subsec:cross_result} investigates how accurately \sys~can identify OSS reuse in practice.  	  
\autoref{subsec:acc_others} compares \sys~with \dejavu, motivating the need for code segmentation. 
In \autoref{subsec:scalability}, we evaluate the scalability of \sys~and the efficacy of redundancy elimination. 
Finally, we introduce our findings on OSS reuse patterns in \autoref{subsec:observation}.
We evaluated \sys on a machine 
with Ubuntu 16.04, 
Intel Xeon Processor @ 2.40 GHz, 32GB RAM, and 6TB HDD.

\medskip
\subsection{Accuracy of \sys}\label{subsec:cross_result}
\PP{Methodology}
We conducted a cross-comparison experiment on our dataset of 10,241 repositories. 
To do so, we first selected representative versions (\ie, the version with the most functions)
of  each OSS. %
As the reused components are mostly similar across different versions in one OSS,
we decided to identify the components only for the representative version for each OSS, 
and measure the detection accuracy.
To evaluate the accuracy of \sys,
we used five metrics:
true positives (TP),
false positives (FP),
false negatives (FN),
precision $\big(\small\frac{\text{\#TP}}{\text{\#TP + \#FP}}\big)$,
and recall $\big(\small\frac{\text{\#TP}}{\text{\#TP + \#FN}}\big)$.

\medskip
\PP{Ground-truth establishment}
Since C/C++ software does not carry standardized information about its components, 
we have to set the criteria for determining whether a detected component is actually reused in the target software.
Therefore, we decided to utilize the following three factors	to verify the detection results:

\smallskip
\begin{center}
	\noindent\fbox{%
		\parbox{0.97\linewidth}{%
			\smallskip
			\begin{itemize}[leftmargin=1.3em]
				\small
				\setlength\itemsep{0.9em}
				\item\textbf{Paths:}
				The file paths of the reused functions
				(for stricter validation,
				we only consider the case
				when the name of the detected component is included in the reused function path);
				\item\textbf{Header files:}
				The header file configured with the OSS name;
				\item\textbf{Metadata files:}
				One of the \texttt{README}, \texttt{LICENSE}, and \texttt{COPYING} files in the top-level directory of the OSS (\cite{kapitsaki2015insight, ikeda2019empirical}). 
			\end{itemize}
			\smallskip
		}%
	}
\end{center}
\smallskip
If one of the above factors of the detected OSS is contained in the target software,
we determine that the detected OSS is the correct component of the target software.
As an example of the paths,
``\texttt{inflate.c}'' of Zlib is reused in the path of ``\path{src/../zlib-1.2.11/inflate.c}''
in MongoDB.
As examples of other factors,
Redis reuses Lua while containing ``\texttt{Lua.h},''	
and
Libjpeg is reused in ReactOS
where the \texttt{README} file of Libjpeg is contained in ReactOS
with the path of
``\path{dll/3rdparty/libjpeg/README}.''

When a false alarm occurs,
neither the name of the falsely detected component is included in the reused function path,
nor the main header file and metadata files of the detected component are reused in the target software.
Moreover,
these factors are only used to verify the results detected by \sys.
Obviously, finding the correct answers is a more complex problem
than verifying the obtained answers,
and an issue arises when identifying components by relying sorely on these factors:
the target software can implicitly reuse an OSS without utilizing both the header files
and the metadata files of the OSS.
Thus,
the validation methods using these factors do not negate the need for \sys.

\PP{Multi-level verification}
Using the aforementioned three factors
(\ie, paths, header files, and metadata files),
we run multi-level verification on the results of \sys:
\begin{enumerate}
	\item \textbf{Automated verification.} We first check that at least one of the three factors of the detected OSS are intact in the target software; this task is done in an automated way.
	\item \textbf{Manual verification.} For the remaining results that are not verified using the automated way, we manually analyzed the results, because the three factors could be reused with modification; for example, OpenSSL could be reused in the path of "\texttt{open\_ssl/}."
	For more accurate verification, we further check whether the name of the identified OSS is specified in any comments of the reused source files in the target software.	
\end{enumerate}
Any identified components that are not verified by the multi-level verification are counted towards FPs.
\begin{figure}[t]
	\begin{flushright}
		\scriptsize{* measured by only using the three automated validation methods.}				
	\end{flushright}
	\begin{center}
		\includegraphics[width=1\linewidth]{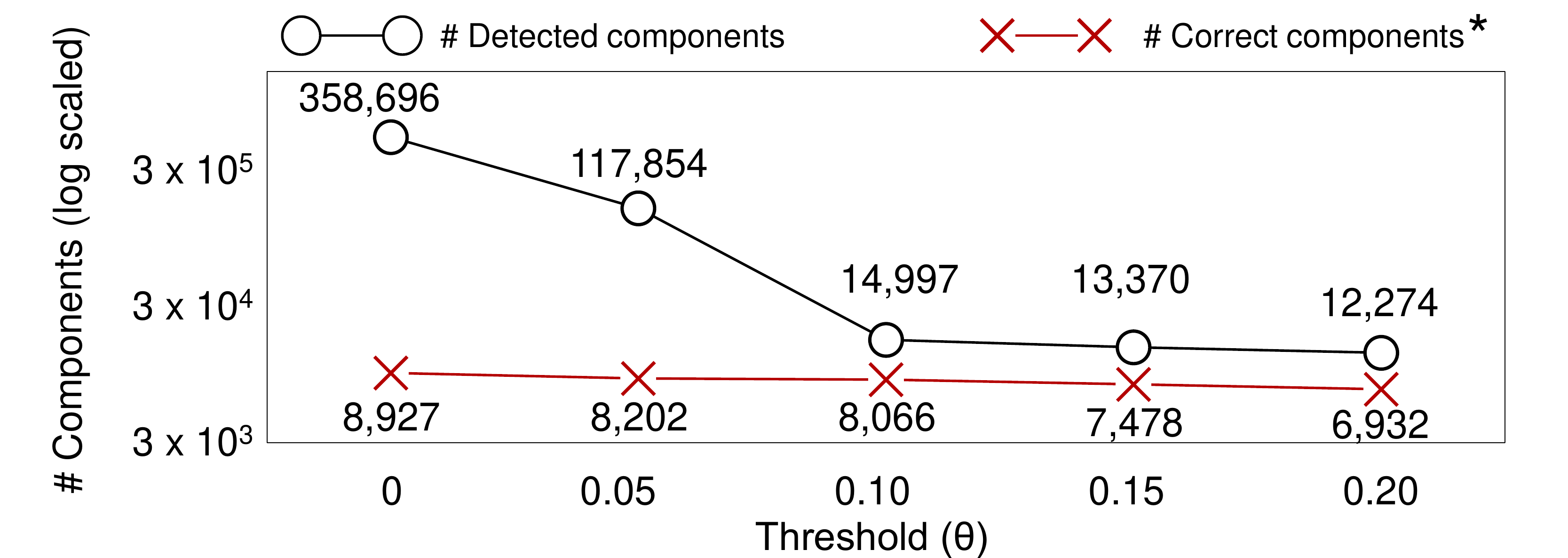}
		\caption{\label{fig:theta_selection} Experimental results for measuring efficiency of $\theta$.}
	\end{center}
	\vspace{-0.7em}
\end{figure}

\PP{Parameter setup} 
We then selected a suitable $\theta$ value
to mitigate false alarms due to widely utilized code
(see \autoref{subsec:p2}).
To select $\theta$,
we evaluated each cross-comparison result
using the predefined automated verification
while setting $\theta$ to 0, 0.05, 0.1, 0.15, and 0.2.
The results are depicted in \autoref{fig:theta_selection}.
Notably,
the proportion of correct components in the detected components
drops significantly
when $\theta$ is less than 0.1.
On the contrary,
if $\theta$ is greater than 0.1,
the proportion of correct components in the detected components
increases slightly;
however,
the number of correct components
decreases.
The overall result implies that a widely utilized code is often shared among different OSS projects
and 
accounts for only a small portion of each OSS project
(generally less than 10\%).
For our experiment,
we set $\theta$ as 0.1
to balance recall and precision.

\PP{Accuracy measurement}
From the cross-comparison result, 
we observed that 4,434 (44\%) out of 10,241 OSS projects 
were reusing at least one other OSS;
a total of 14,997
components
were detected.
As it is challenging to identify literally every component
in the target software,
we cannot easily measure false negatives.
Hence, we only considered false negatives that occurred when the application code of an OSS 
is reused less than the $\theta$ ratio
and thus \sys fails to identify it,
which can be measured by
subtracting the number of correct components when $\theta$ is 0.1
from that when $\theta$ is zero
(see \autoref{fig:theta_selection}).
Among the cross-comparison results,
we successfully validated 8,066 results (54\%)
using the \textit{automated} verification,
and the remaining 6,931 detection results
were analyzed by the \textit{manual} verification.
The manual verification was performed by two people and
took two weeks.
We manually viewed the paths,
header files, and metadata files,
as well as the reused source code and comments within the source code %
to determine whether the identified OSS is the correct component.
The accuracy measurement results are presented in \autoref{table:acc_res}.

Although most of the detected OSS components were reused with modification (95\%),
\sys achieved \textbf{91\% precision} 
and
\textbf{94\% recall}.
Although \sys precisely identified reused components in most cases,
it reported several false results.
We observed that false positives were mainly caused when the target software and the detected component
only shared the third-party software that was not included in our component DB.
Hence, the application code of OSS projects was not properly obtained, resulting in false alarms.
In addition,
if the reuse ratio of the application code of the OSS was less than $\theta$,
or the reused component was not included in the component DB, \sys failed to detect the correct OSS components,
\ie, false negatives occurred.	
However, simply decreasing $\theta$ for reducing false negatives can impair precision.
Expanding current component DB such as collecting more OSS projects from various sources
would be an efficient solution to reduce false results,	
even so,
we believe that the method of minimizing false alarms through the proposed code segmentation works efficiently,
and the
selected $\theta$ simultaneously maintains a good balance in terms of precision and recall.

\begin{table}[t]
	\footnotesize
	\renewcommand{\tabcolsep}{2.1mm}
	\begin{center}
		\caption{\label{table:acc_res}%
			Accuracy of \sys component identification results.
		}
		\vspace{-0.8em}
		\begin{tabular}{rccccc}
			\thickhline
			\multirow{2}{*}{\begin{tabular}[c]{@{}c@{}}\rule{0in}{2ex}\\\textbf{(For 14,997 cases)}\end{tabular}}
			& \multicolumn{5}{c}
			{\begin{tabular}[c]{@{}c@{}}\rule{0in}{2.5ex}\textbf{Validation result}\end{tabular}}\\\cline{2-6}
			\rule{0in}{2.5ex}
			& \textbf{\#\textsc{Tp}}
			& \textbf{\#\textsc{Fp}}
			& \textbf{\#\textsc{Fn}}
			& \textbf{Precision}
			& \textbf{Recall}
			\\\hline
			\multicolumn{6}{l}
			{\begin{tabular}[l]{@{}c@{}}\rule{0in}{2.2ex}
					{\textit{Automated verification results}}\end{tabular}}\\
			\rule{0in}{2.5ex}Paths ($V_P$)
			&3,685
			& \scriptsize{\textcolor{gray}{\texttt{N/A}}}
			& \scriptsize{\textcolor{gray}{\texttt{N/A}}}
			& \scriptsize{\textcolor{gray}{\texttt{N/A}}}
			& \scriptsize{\textcolor{gray}{\texttt{N/A}}}
			\\
			\rule{0in}{1ex}Header files ($V_H$)
			&3,286
			& \scriptsize{\textcolor{gray}{\texttt{N/A}}}
			& \scriptsize{\textcolor{gray}{\texttt{N/A}}}
			& \scriptsize{\textcolor{gray}{\texttt{N/A}}}
			& \scriptsize{\textcolor{gray}{\texttt{N/A}}}
			\\
			\rule{0in}{1ex}Metadata files ($V_M$)
			&4,175
			& \scriptsize{\textcolor{gray}{\texttt{N/A}}}
			& \scriptsize{\textcolor{gray}{\texttt{N/A}}}
			& \scriptsize{\textcolor{gray}{\texttt{N/A}}}
			& \scriptsize{\textcolor{gray}{\texttt{N/A}}}
			\\\hline
			\multicolumn{6}{l}
			{\begin{tabular}[l]{@{}c@{}}\rule{0in}{2.4ex}
					{\textit{Combined all automated verification methods ($V_P \cup V_H \cup V_M$)}}\end{tabular}}\\
			\rule{0in}{2.5ex}
			& 8,066 %
			& \scriptsize{\textcolor{gray}{\texttt{N/A}}}
			& \scriptsize{\textcolor{gray}{\texttt{N/A}}}
			& \scriptsize{\textcolor{gray}{\texttt{N/A}}}
			& \scriptsize{\textcolor{gray}{\texttt{N/A}}}
			\\\hline
			\multicolumn{1}{l}{\rule{0in}{2.4ex}{\textit{  Manual verification}}}
			&5,510
			&1,421
			&861 
			& 0.80
			& 0.86
			\\\hline
			\multicolumn{1}{l}{\rule{0in}{2.4ex}\textbf{\textit{  Total}}}
			& \textbf{13,576}
			& \textbf{1,421}
			& \textbf{861}
			& \textbf{\color{tomato}{0.91}}
			& \textbf{\color{tomato}{0.94}}
			\\\thickhline				
		\end{tabular}
	\end{center}
	
	\vspace{0.3em}
	{\footnotesize *According to our definition, all the results verified by the validation methods are TP, thus, the remaining columns are filled with \scriptsize{\textcolor{gray}{\texttt{N/A}}}.}
	\vspace{-0.8em}
\end{table}	

\PP{Version identification accuracy}
Some components are not managed by the versioning system, 
and further
the target software often does not reuse files or codes containing version information of a reused component.
Subsequently,
we decided to measure version identification accuracy for
the three most widely reused OSS projects in our results:
GoogleTest, Lua, and Zlib. %
Their version information
is relatively well-defined compared to that of other OSS %
while still providing a sufficient pool to measure the accuracy.	
In our cross-comparison experiment,
these three OSS projects were reused a total of 682 times.
Approximately half of the reuses provided the utilized version %
using related files: ``zlib.h'' in Zlib, \texttt{README} in Lua, and
\texttt{CHANGES} in GoogleTest.
When these files were not reused in the target software,
the version information was manually analyzed (\eg, using the commit log). 
The version identification result is presented in \autoref{table:version}.
Partial reuse mainly occurred in Lua,
code changes mostly appeared in Zlib,
and structural changes primarily arose in GoogleTest.

\sys succeeded in identifying
the utilized version information
with 91.5\% precision.
We failed to identify the accurate version in some modified reuse cases,
especially when the functions from different versions
(in extreme cases, more than 10 versions)
of an OSS were mixed in the target software,
\ie, code-changed reuse.
In such cases,
we determined that not only is identifying the correct version a challenge,
but also that the version identification is meaningless.
Therefore, 
we concluded that identifying the OSS reuse and the most similar version
would be sufficient for the code-changed reuses.

\medskip
\subsection{In-depth comparison with \dejavu}\label{subsec:acc_others}
\PP{Tool selection}
We reviewed several related approaches
published since 2010; %
however, most SCA approaches
are only applicable to identifying components in Android applications
or software binaries
(\cite{ma2016libradar, backes2016reliable, duan2017identifying, li2017libd, tang2018bcfinder, copilot}).
For example,
OSSPolice \cite{duan2017identifying}
is open to the public
but its targets are Android applications. 
Moreover, as the parser for the C/C++ library is not open source,
it would be difficult to apply their algorithm to our experiment.
Therefore,
we decided to compare \sys with \dejavu,
a similar approach 
in terms of technology and purpose
\cite{lopes2017dejavu}.
\dejavu is based on the code clone detection technique (\ie, SourcererCC \cite{sajnani2016sourcerercc}),
and it aims to analyze the software dependencies among GitHub repositories
by detecting project-level clones.
Thus, we concluded that the detection results of \dejavu can be compared with those of \sys.

\begin{table}[t]
	\footnotesize
	\renewcommand{\tabcolsep}{4.8mm}
	\begin{center}
		\caption{\label{table:version} Version identification accuracy of \sys. %
		}
		\vspace{-0.8em}
		\begin{tabular}{crrr}
			\thickhline
			\multicolumn{1}{c}{\rule{0in}{2.2ex}\textbf{Reuse patterns}} 
			& \textbf{\#\textsc{Tp}} & \textbf{\#\textsc{Fp}} & \textbf{Precision} \\\hline
			\multicolumn{1}{l}{\rule{0in}{2.2ex}{\textit{Exact reuse}}} E
			& 115
			& 0
			& 100\% \\\hline
			\multicolumn{1}{l}{\rule{0in}{2.2ex}{\textit{Modified reuse}}} P
			& 112& 3&97\%\\			%
			\multicolumn{1}{r}{\rule{0in}{1ex}\hspace{3em} } SC   & 25   & 0  & 100\% \\ 
			\multicolumn{1}{r}{\rule{0in}{1ex}\hspace{3em} } P \& CC   & 185  & 29 & 86\% \\ 
			\multicolumn{1}{r}{\rule{0in}{1ex}\hspace{3em} } P \& SC \& CC & 187  & 26 & 88\% \\ \hline
			\multicolumn{1}{l}{\rule{0in}{2.2ex}\textbf{\textit{Total}}}
			& 624  & 58 & \textcolor{tomato}{\textbf{91.5\%}} \\ \thickhline
		\end{tabular}
		
		\scriptsize{\vspace{0.3em} E: Exact reuse, P: Partial reuse, SC: Structure-Changed reuse, CC: Code-Changed reuse}
		
		\vspace{-2.5em}
	\end{center}
\end{table}

\PP{Methodology}
Currently,
the \dejavu software is not publicly available;
only the detection results
previously obtained using the dataset
(\ie, GitHub C/C++ repositories in 2016)
are provided\footnote{\url{http://mondego.ics.uci.edu/projects/dejavu/}}.
We thus attempted to examine 
the component identification results 
of the common datasets between 
\sys and \dejavu,
and compare them.
In particular,
\dejavu determined the existence of a dependency based on the code similarity score
between the two software projects.
We set the similarity threshold to 50\%, 80\%, and 100\% in \dejavu
(refer to \cite{lopes2017dejavu})
and analyzed the number of correct target software and OSS component pairs from their results where the similarity score exceeded the selected threshold.
\sys employs $\theta$ as 10\%.
To demonstrate the efficiency of code segmentation,
we provide both component identification results
when code segmentation is turned on and off.

\PP{Comparison results}
Among our cross-comparison results,
four %
of the top 50 software projects (\ie, ArangoDB, Crown, Cocos2dx-classical, and Splayer)
with the maximum OSS reuse were observed to be part of the \dejavu datasets as well.
We decided to compare the OSS component detection results 
between \sys and \dejavu
for these four software projects;
the results are listed in \autoref{table:acc_dejavu}.

\dejavu
failed to identify many modified components.
In fact,
most identified components for the selected software
were reused with modifications (see \autoref{table:modified_dejavures}).
As \dejavu could not identify components
when the reused code ratio was less than the selected threshold,
the results showed low recall values (\ie, at most 40\%).
Moreover, although \dejavu aimed to detect project-level clones, 
its mechanism did not include a handling routine for false positives caused by nested OSS.
Subsequently,
\dejavu reported 
many false positives,
\ie, it showed 4\% and 7\% precision when the threshold was selected as 50\% and 80\%, respectively.	
Even though \dejavu showed
100\% precision when the threshold was selected as 100\%,
it could not detect any partially reused components,
as indicated by the fact that the recall was 16\%.

\begin{table}[t]
	\scriptsize
	\renewcommand{\tabcolsep}{0.5mm}
	\begin{center}
		\caption{\label{table:acc_dejavu}Component identification results of \dejavu and \sys.}
		\vspace{-1em}
		\begin{tabular}{crrr|rrr|rrr|rrr|rrr}			
			\thickhline
			\multicolumn{1}{c}{} 
			&\multicolumn{3}{c}{\multirow{2}{*}{\begin{tabular}[c]{@{}c@{}}\rule{0in}{2.5ex}\textbf{\sys}\\{(with $cs$)}\end{tabular}}}
			&\multicolumn{3}{c}{\multirow{2}{*}{\begin{tabular}[c]{@{}c@{}}\rule{0in}{2.5ex}\textbf{\sys}\\{(without $cs$)}\end{tabular}}}
			&\multicolumn{9}{c}{\rule{0in}{2ex}\textbf{\dejavu} (classified by the threshold)}\\\cline{8-16}
			& 
			&
			& \multicolumn{1}{c}{} 
			&
			&
			& \multicolumn{1}{c}{} 
			& \multicolumn{3}{c}{\rule{0in}{2.2ex}\textbf{50\%}} 
			& \multicolumn{3}{c}{\textbf{80\%}} 
			& \multicolumn{3}{c}{\textbf{100\%}} \\

			\multicolumn{1}{c|}{\rule{0in}{2.5ex}\textbf{Software}}
			& \multicolumn{1}{r}{\textbf{\#T}} 
			& \multicolumn{1}{r}{\textbf{\#\textsc{Fp}}} 
			& \multicolumn{1}{r|}{\textbf{\#\textsc{Fn}}} 
			& \multicolumn{1}{r}{\textbf{\#T}} 
			& \multicolumn{1}{r}{\textbf{\#\textsc{Fp}}} 
			& \multicolumn{1}{r|}{\textbf{\#\textsc{Fn}}} 
			& \multicolumn{1}{r}{\textbf{\#T}} 
			& \multicolumn{1}{r}{\textbf{\#\textsc{Fp}}} 
			& \multicolumn{1}{r|}{\textbf{\#\textsc{Fn}}} 
			& \multicolumn{1}{r}{\textbf{\#T}} 
			& \multicolumn{1}{r}{\textbf{\#\textsc{Fp}}} 
			& \multicolumn{1}{r|}{\textbf{\#\textsc{Fn}}} 
			& \multicolumn{1}{r}{\textbf{\#T}} 
			& \multicolumn{1}{r}{\textbf{\#\textsc{Fp}}} 
			& \multicolumn{1}{r}{\textbf{\#\textsc{Fn}}} \\\hline
			\multicolumn{1}{r|}{\rule{0in}{2.2ex}\textbf{\textit{ArangoDB}}}
			& 29 & 2 & 0 
			& 29 & 450 & 0 
			& 11 & 411 & 18 
			& 8 & 236 & 21 
			& 7 & 0 & 22\\
			\multicolumn{1}{r|}{\rule{0in}{2.2ex}\textbf{\textit{Crown}}}
			& 23 & 0 & 0
			& 23 & 750 & 0
			& 9 & 171 & 14
			& 6 & 23 & 17
			& 3 & 0 & 20\\
			\multicolumn{1}{r|}{\rule{0in}{2.2ex}\textbf{\textit{Cocos2dx}}}
			& 19 & 2 & 0
			& 19 & 231 & 0
			& 8 & 52 & 11
			& 2 & 6 & 17 
			& 1 & 0 & 18 \\
			\multicolumn{1}{r|}{\rule{0in}{2.2ex}\textbf{\textit{Splayer}}}
			& 16 & 1 & 0
			& 16 & 275 & 0
			& 7 & 236 & 9
			& 6 & 27  & 10
			& 3 & 0 & 13 \\\hline
			\multicolumn{1}{r|}{\rule{0in}{2.2ex}\textbf{\textit{Total}}}
			&{\textcolor{tomato}{\textbf{87}}}&\textcolor{tomato}{\textbf{5}} &\textcolor{tomato}{\textbf{0}} 
			&{87}&{1,706}&{0}
			&{35}&{870}&{52}
			&{22}&{292}&{65}
			&{14}&{0}&{73}\\\hline
			\multicolumn{1}{r|}{\rule{0in}{2.2ex}\textbf{\textit{Precision}}}
			&\multicolumn{3}{c|}{\textbf{\textcolor{tomato}{0.95}}}
			&\multicolumn{3}{c|}{\textbf{0.05}}
			&\multicolumn{3}{c|}{\textbf{0.04}}
			&\multicolumn{3}{c|}{\textbf{0.07}}
			&\multicolumn{3}{c}{\textbf{1.0}}\\\hline
			\multicolumn{1}{r|}{\rule{0in}{2.2ex}\textbf{\textit{Recall}}}
			&\multicolumn{3}{c|}{\textbf{\textcolor{tomato}{1.0}}}
			&\multicolumn{3}{c|}{\textbf{1.0}}
			&\multicolumn{3}{c|}{\textbf{0.40}}
			&\multicolumn{3}{c|}{\textbf{0.25}}
			&\multicolumn{3}{c}{\textbf{0.16}}\\
			\thickhline
		\end{tabular}

		\scriptsize{\vspace{0.3em} $cs$: code segmentation; \#T: the number of true positives;\\\#\textsc{Fp}: the number of false positives; \#\textsc{Fn}: the number of false negatives.}
	\end{center}
	
	\vspace{1.0em}		
\end{table}

\begin{table}[t]
	
	\renewcommand{\tabcolsep}{2.2mm}
	\begin{center}
		\caption{\label{table:modified_dejavures}OSS reuse patterns in the four software projects.}
		\vspace{-1em}
		\begin{tabular}{rcccc}
			\thickhline
			\multirow{2}{*}{\begin{tabular}[c]{@{}c@{}}\rule{0in}{2.5ex}\\\textbf{Software}\end{tabular}}
			&\multicolumn{4}{c}{\rule{0in}{2ex}\textbf{Reuse patterns}}\\
			\cline{2-5}
			&\rule{0in}{2.2ex}\textbf{E}xact
			& \textbf{P}artial
			& \textbf{S}tructure-\textbf{C}hanged
			& \textbf{C}ode-\textbf{C}hanged\\\hline
			\rule{0in}{2ex}\textbf{\textit{ArangoDB}}
			&7                                                                               &19                                                        
			&4                                                                   
			&11\\                                   
			\rule{0in}{1.7ex}\textbf{\textit{Crown}}
			&3                               
			&20                                                         
			&3                                                                    
			&16\\
			\rule{0in}{1.7ex}\textbf{\textit{Cocos2dx}}
			&1                                                                               &17                                                         
			&2                                                                   
			&14\\
			\rule{0in}{1.7ex}\textbf{\textit{Splayer}}
			&3                                                                               
			&10 
			&4                                                                    
			&7
			\\\hline
			\rule{0in}{2ex}\textbf{\textit{Total}}
			&\textbf{14}
			&\textbf{\color{tomato}{66}}
			&\textbf{\color{tomato}{13}}
			&\textbf{\color{tomato}{48}}
			\\
			\thickhline
		\end{tabular}

		\scriptsize{\vspace{0.3em} *P, SC, and CC can occur simultaneously in the modified component.}			
		\vspace{-1.5em}
	\end{center}
\end{table}

In contrast, 
\sys yielded substantially 
better accuracy than \dejavu,
\ie, 95\% precision and 100\% recall
when the code segmentation is applied.
In the absence of code segmentation,
\sys reported numerous false positives (\ie, 5\% precision)
with the same cause as \dejavu:
this implies that the method of using only the application code of OSS
for matching through code segmentation can successfully filter out countless false positives.
Lastly,
OSS components that were identified only in \dejavu
and not identified in \sys
did not appear in the four software projects.

\smallskip
\subsection{Speed and Scalability of \sys}\label{subsec:scalability}
\PP{Efficacy of redundancy elimination}
We can reduce space complexity
by eliminating redundancies across OSS versions.
The total number of functions
in all versions of every OSS project that we collected
is 2,205,896,465.
After eliminating redundancies,
we confirm that the number of non-redundant functions only accounts for 
2.2\% (49,330,494 functions) of the total functions,
indicating that the size of the comparison space can be reduced by 45 times
compared to all functions.

\PP{Speed}
When we measured the preprocessing (extracting functions from the OSS, storing hashed functions, and generating the component DB) time of \sys, on average, it took 1 min to preprocess 1 million LoC.
Note that the OSS does not need to be preprocessed 
again after it undergoes initial preprocessing. 
In contrast,
component identification occurs frequently.
Hence,
it is necessary to achieve fast speeds 
for practical use.
When we compared 10,241 representative versions with the component DB, %
\sys took less than 100 hours in total. %
This implies that
\sys takes less than a minute per target application on average,
which is sufficiently fast for practical use.

\begin{figure}[t]
	\begin{center}
		\includegraphics[width=1.0\linewidth]{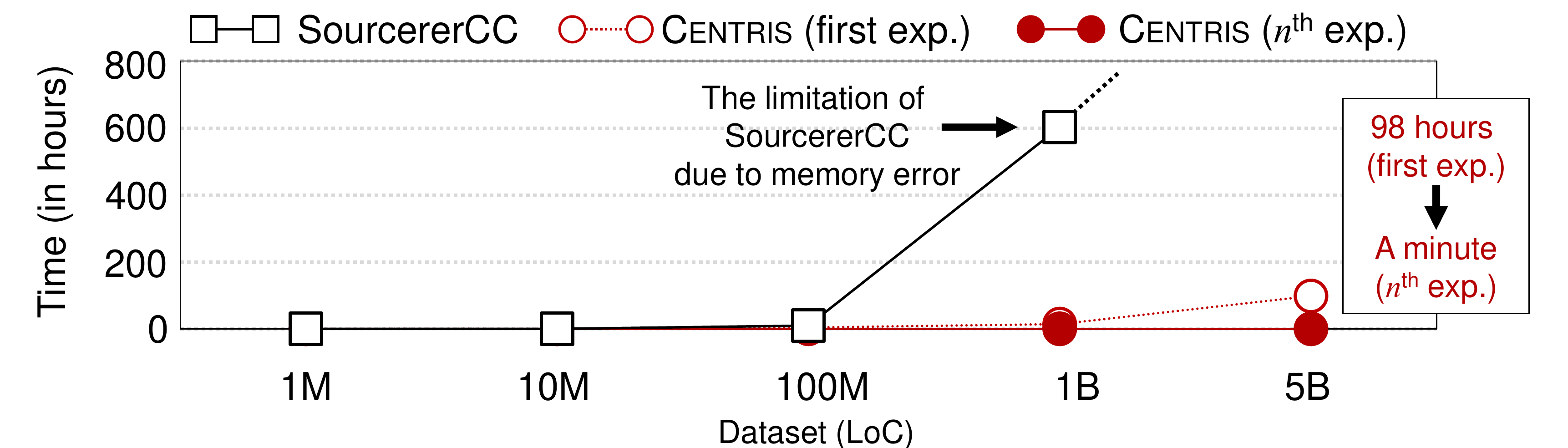}
		\vspace{-0.8em}			
		\caption{\label{fig:scal} Total time consumed on varying dataset sizes.
			\sys exhibited tremendous time efficiency because of recycling of the preprocessed OSS projects in the dataset,
			whereas SourcererCC required approximately three weeks to process the matching using the 1 billion LoC dataset.
		}
	\end{center}

	\vspace{0.4em}
\end{figure}	

\PP{Scalability}
To evaluate the scalability of \sys, 
we measured the time taken to compare
the target software of 1 million LoC
with different datasets
ranging from 1 million to 5 billion LoC. 
We compared the performance of \sys 
with that of the core algorithm of \dejavu
(SourcererCC \cite{sajnani2016sourcerercc}).
\autoref{fig:scal} depicts 
the results.
In the first experiment, 
\sys required 98 hours
for preprocessing and matching. 
After the first experiment,
because \sys could recycle the 
preprocessed component DB,
the required time was significantly reduced
to less than a minute.
SourcererCC needed three weeks for 1 billion LoC dataset, 
and when the dataset was increased in size,
we could not measure the time consumed owing to memory errors
in our evaluation environment;
even if the experiment is performed with a larger memory, 
we expect the processing time to be significantly high.

\subsection{Findings on OSS reuse patterns}\label{subsec:observation}
\begin{figure}[t]
	\begin{center}
		\includegraphics[width=1\linewidth]{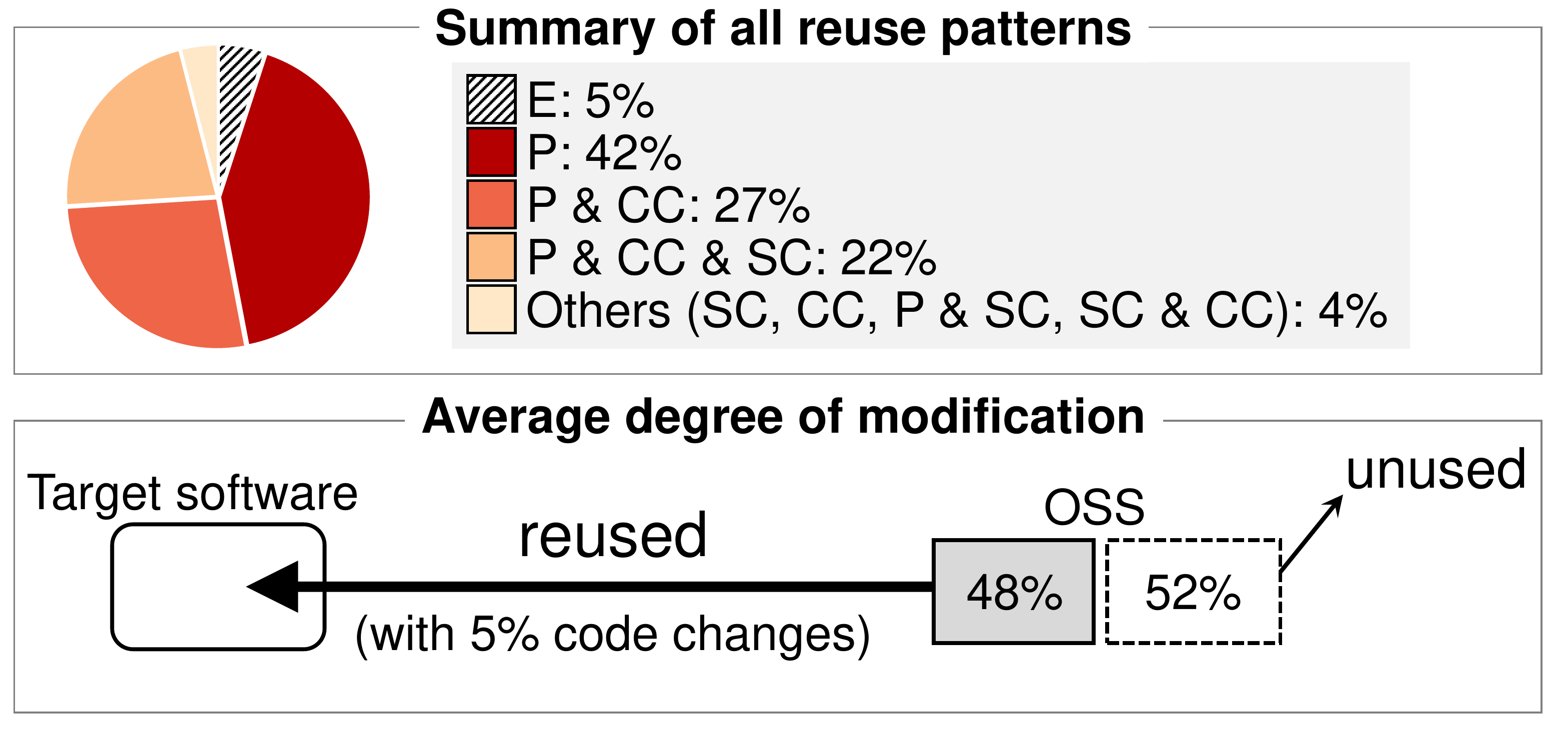}
		\vspace{0.1em}
		\caption{\label{fig:observe}Depiction of detected reuse patterns
			and averaged modification degree obtained from our experiment.
			Partial reuse appeared the most, and we found that developers reused only half of the OSS code base with 5\% code changes on average.}
	\end{center}
	
	\vspace{0.5em}
\end{figure}

From the cross-comparison result,
we found that 4,434 (44\%) OSS projects were reusing at least one other OSS.
Surprisingly,
the \textit{modified reuses accounted for 95\%}
of the detected components.
The distribution of detected reuse patterns and the average degree of modification are depicted in \autoref{fig:observe}.
We summarize two key observations as follows.

\PP{Partial reuse accounts for 97\% of all modified reuses}
We observed that
developers were mostly reusing only part of the OSS code base they needed.
Mainly,
functions deemed unnecessary for the target software to perform the desired operation,
testing-purposed functions such as located in ``\texttt{test}/'' directory,
and cross-platform support functions were excluded during an OSS reuse.

\PP{Code and structural changes also frequently occur}
Among all modified reuses,
53\% changed at least one original function,
and
26\% changed the original structure.
We found that
code changes occurred primarily due to developers' attempts
to adapt the reused functions to their software
(\eg, change variable names),
and to fix software vulnerabilities propagated from reused OSS.
Moreover,
we observed that the reused functions of an OSS are often merged in a single file
rather than scattered in different structures.
For example, Rebol software
reused only 30\% of Libjpeg while integrating all of the
reused functions into ``\path{src/core/u-jpg.c}.''

Our observation results suggest the need to detect heavily modified components
(\ie, only 48\% of the OSS code base were reused on average),
but the existing approaches did not consider this trend
(\eg, both \dejavu and OSSPolice selected the lowest threshold as 50\%),	
hence failed to identify many correct components.
From this point of view,
\sys would be a better solution for the efficient SCA process,
as it can precisely identify modified components.	

\medskip
\section{Discussion}\label{sec:dis}

\vspace{0.6em}
\subsection{Function-level granularity}
\vspace{0.3em}

Although the design of \sys is applicable to any granularity,
the benefits we can obtain by using each granularity are certainly different 
in terms of both the accuracy and the scalability in component identification.
We confirmed that the function-level granularity works best for balancing scalability and accuracy in component identification, thus, the function units were used for our experiments as the basis.

If \sys uses a coarse-grained unit, \eg, a file, \sys is able to identify components with higher scalability, however, \sys misses many partial and code-changed reuses.
To demonstrate this,
we analyzed the 14,997 components detected by \sys in \autoref{sec:eval}.
Specifically, when we detected cases where more than 10\% (\ie, $\theta$) of all files in the component were reused exactly,
only 60\% (8,992 cases) of them belonged to these cases.
This is because developers often reuse only necessary functions in a file while excluding unnecessary functions
(\eg, functions used for testing).

Conversely,
if \sys uses a finer-grained unit, \eg, a line or a token,
\sys can analyze more detailed reuse patterns, yet the disadvantages are clear:
(1) the poor scalability and (2) more false alarms due to a short, generic code.
Our component DB contains a total of 80 billion lines of source code,
and it is not trivial to compare them with all the lines of source code of the target software.
In addition, as simple and generic codes, \eg, a variable declaration code line such as ``\texttt{int i;}'', are widely spread among software that does not have a reuse relation, this yields more false alarms.

Thus, we determined that the function-level granularity was most balanced:
reasonable scalability (see \autoref{subsec:scalability}),
fewer false positives and false negatives than the line-level and file-level granularity,
respectively
(the benefits of the function-level granularity have been discussed in previous studies~\cite{sajnani2016sourcerercc, kim2017vuddy, duan2017identifying, duan2019automating},
specifically,
VUDDY~\cite{kim2017vuddy} introduced the scalability and accuracy comparisons between function-units and other units in detail).

\vspace{0.5em}
\subsection{Generalizability of \sys}
\vspace{0.3em}

The generalizability of a tool is an important issue from a practical point of view~\cite{ghaisas2013generalizing, wieringa2015six}.
In \autoref{sec:eval},
we evaluated \sys over 10,241 extensively-reused popular GitHub projects,
considering them as a representative body of OSS, and observed promising results.
This gives us confidence that \sys will work well in all contexts of OSS projects that fit in the ecosystem.

For one thing, the code segmentation of \sys is affected by the number of OSS contained in the dataset (\ie, the component DB).
If the approach of \sys performed with fewer OSS than used in this paper,
the identification accuracy may slightly decrease;
meanwhile, if \sys identifies components with a larger and more refined dataset,
the higher component identification accuracy can be obtained.
We leave the task of finding the most optimal number of OSS that will be contained in the component DB to future work.

\vspace{0.5em}
\subsection{Implications for practice}
\vspace{0.3em}
To the best of our knowledge,
none of the existing approaches are applicable to precisely identify modified OSS components.
Yet, our experimental results affirmed that the modified reuses are prevalent in the real-world popular OSS ecosystem.

\sys is a design science research~\cite{wieringa2014design} with a clear goal to design and improve an algorithm (\ie, artifact) in the context of identifying modified OSS reuse from the software,
based on two techniques: code segmentation and redundancy elimination.
From this point of view, \sys can be the first step towards addressing problems arising from unmanaged OSS components in practice.
In particular,
with the help of \sys, developers can precisely identify the reused modified components,
and further address potential threats arising from unmanaged components (\eg, by updating components).

\vspace{0.6em}
\subsection{Use case: software vulnerability management}\label{sec:usecase}
\vspace{0.3em}
One potential use case of \sys is software vulnerability management, which reduces security issues by identifying newly found but unpatched vulnerabilities.   
Below, we discuss our experience of using \sys in this regard. 

By referring to the National Vulnerability Database (NVD),
we can obtain the affected software and version information,
\ie, Common Platform Enumeration (CPE),
for each reported vulnerability.
We have extensively examined whether the names and versions of detected OSS components
are included in the obtained CPE~\cite{duan2017identifying}.
Consequently,
\sys discovered that 572 OSS projects
contain at least one other vulnerable OSS component.
Among them,
27 OSS projects
are still reusing the vulnerable OSS in their latest version.

For the cases of successfully reproducing the vulnerability,
we have reported to the corresponding vendors.
Of these, the most notable example related to modified reuse is Godot (32K GitHub stars).
We found that the latest version of Godot %
was reusing vulnerable JPEG-compressor contains CVE-2017-0700 (CVSS 7.8).
Godot was reusing only \textit{one file} from JPEG-compressor (``\texttt{jpgd.cpp}''),
which contains the exact vulnerable code.
More seriously, this vulnerability could be reproduced by
simply uploading a malicious image file to the Godot project.	
We reported this information on their repository’s issues;
developers immediately patched the vulnerability (Jul. 2019).

Likewise,
we could successfully reproduce a vulnerability in Stepmania, Audacity (so far, reusing vulnerable Libvorbis),
LibGDX (reusing vulnerable JPEG-compressor),
and Redis (reusing vulnerable Lua);
for all cases, we reported to the corresponding development and security teams,
and confirmed that proper actions were taken, such as vulnerability patches.
Even though developers reuse only a small part of an OSS,
the vulnerability in that part
opens up the attack surface.
To address this,
we can apply \sys for more attentive vulnerability management as shown here.

\vspace{0.6em}
\subsection{Threats to validity}
\vspace{0.3em}

First,
although our dataset is more expansive compared to those in previous approaches,
the benchmark OSS projects utilized herein %
might not be representative.
Second,
to the best of our knowledge,
there are no approaches that directly attempt to identify modified components.
Although we conducted an in-depth comparison with \dejavu,
our intention is not to deny the accuracy and performance of \dejavu,
but to demonstrate that our approach is much more efficient
for the purpose of identifying modified components.
Finally,
there may be hidden components in a target software
that both \sys and \dejavu
failed to identify;
as all OSS reuse statuses are not known,
we cannot exactly measure the missed components,
and these are the false negatives of \sys.

\medskip
\section{Related Work}\label{sec:related}
\PP{Code clone detection}
Over the past decades, numerous techniques
have been
proposed to detect code clones~\cite{baker1995finding, baxter1998clone, komondoor2001using, kamiya2002ccfinder, myles2004detecting, li2004cp, myles2005k, jiang2007deckard, schleimer2003winnowing, roy2007survey, roy2008nicad,  sajnani2016sourcerercc, semura2017ccfindersw, lopes2017dejavu, kim2017vuddy, nishi2018scalable, searchcode, wang2018ccaligner, luciv2018duplicate, gharehyazie2018cross, vislavski2018licca},
and \sys adopts a signature-based clone detection method~\cite{kim2017vuddy}. 
However, as we demonstrated in this paper, using an existing clone detection technique as it is suffers from false alarms when identifying modified reuse of nested OSS. 

\PP{Software composition analysis}
Existing SCA techniques~\cite{ma2016libradar, backes2016reliable, duan2017identifying, li2017libd, tang2018bcfinder, copilot}
are not accurate enough to identify modified OSS reuse. 
Duan et al. \cite{duan2017identifying} proposed OSSPolice
to find third-party libraries of an Android application.
They utilized constant features
of obfuscation
to extract
the version information
and determine
if vulnerable versions were utilized.
They minimized false alarms
through hierarchical indexing and matching. %
Since their concern is more on accurately identifying third-party libraries at the binary level,
thus, it differs from our concern to detecting modified components.
Backes et al. \cite{backes2016reliable}
and Bhoraskar et al. \cite{bhoraskar2014brahmastra} %
also do not consider
detection of modified OSS reuse.
CoPilot \cite{copilot}
analyzes security risks that
arise from unmanaged OSS components.
However, as it is based on dependency files,
it can be applied only for languages
in which dependencies are managed.
To our knowledge and experience, commercial SCA tools such as 
Black Duck Hub \cite{blackduckopenhub} by Synopsys \cite{synopsys}
and Antepedia \cite{antepedia}
do not consider modified reuse
and hence miss
many reused components. 

\medskip
\section{Conclusion}\label{sec:conclusion}
\smallskip

Identifying OSS reuse is a pressing issue in modern software development practice, 
because unmanaged OSS components pose threats by increasing critical security and legal issues. 
In response, we presented \sys, which departs significantly from existing techniques by enabling precise and scalable 
identification of reused OSS components
even when they are heavily modified and nested. 	
With the information provided by \sys,
developers can mitigate threats
that arise from unmanaged OSS components,
which not only increases
the maintainability of the software,
but also
renders a safer development environment.
\section*{Data Availability}
We service \sys as a form of open web service at IoTcube~\cite{kim2017poster} (\url{https://iotcube.net/}).
The source code and dataset (\ie, the component DB used in \autoref{sec:eval})
is available at \url{https://github.com/wooseunghoon/Centris-public}.

\section*{Acknowledgment}
We appreciate the anonymous reviewers for their valuable comments to improve the quality of the paper.
This work was supported by Institute of Information \& Communications Technology Planning \& Evaluation (IITP) grant 
funded by the Korea government (MSIT) (No.2019-0-01697 Development of Automated Vulnerability Discovery Technologies for Blockchain Platform Security and No.2020-0-01819 ICT Creative Consilience program).

\end{document}